**New zero Poisson's ratio models**

Vladimir Gaal,[1] Varlei Rodrigues,[1] Socrates E. Dantas,[2] Douglas S. Galvão,[3] and Alexandre F. Fonseca[3, a)]

[1] Applied Physics Department, University of Campinas, Campinas, SP, 13083-970, Brazil.

[2] Physics Department, Science Institute, Universidade Federal de Juiz de Fora, Juiz de Fora, MG, 36036-330, Brazil.

[3] Applied Physics Department, University of Campinas, Campinas, SP, 13083-970, Brazil.

Most materials exhibit positive Poisson's ratio (PR) values but special structures can also present negative and, even rarer, zero (or close to zero) PR. Null PR structures have received much attention due to their unusual properties and potential applications in different fields, such as aeronautics and bio-engineering. Here, we present a new and simple near-zero PR 2D topological model based on a structural block composed of two smooth and rigid bars connected by a soft membrane or spring. It is not based on re-entrant or honeycomb-like configurations, which have been the basis of many null or quasi-null PR models. Our topological model was 3D printed and the experimentally obtained PR was −0.003 ± 0.001, which is one the closest to zero value ever reported. This topological model can be easily extended to 3D systems and with compression in any direction. The advantages and disadvantages of these models are also addressed.

Formally defined by Poisson in 1827,[1] the so-called Poisson's ratio (PR) is largely used to characterize material mechanical properties.[2] PR is defined as the negative ratio of transverse strain ($\varepsilon_x$) to longitudinal strain ($\varepsilon_z$) along the direction of a stretching or a compression stress $\sigma_z = \varepsilon_x / \varepsilon_z$.[3,4]

The vast majority of materials present positive PR since $\varepsilon_x$ and $\varepsilon_z$ usually have opposite signs, i. e., they expand (contract) crosswise - x direction - when compressed (tensioned) longwise - z direction. A few materials possess negative PR (so-called auxetics),[5–7] i. e., they contract (expand) laterally when compressed (tensioned) longitudinally. Auxeticity has been seen in foams,[5,6,8] crystalline hinged structures,[9,10] nanomaterials,[11] and in certain extreme states of matter.[12] Zero or close to zero PR materials, i. e., materials that neither contract nor expand laterally under the application of tensile or compressive stresses such as, for example, cork[2,13] and entangled springs,[14,15] are even rarer and have received much attention towards applications in different fields as aeronautics[16,17] medicine,[18] and tissue engineering.[19]

One paradigm for the negative PR behavior is the so-called *re-entrant* honeycomb structures, which are structures with inward protruded ribs,[6,20,21] see Fig. 1(a). *Semi-re-entrant* structures, Fig. 1(b), combine regular (positive PR) and re-entrant (negative PR) honeycomb topologies in such a way that they cancel each other and the net effect is a null PR behavior.[22] This has been exploited to create zero or quasi zero PR structures, such as *SILICOMB*,[23,24]

*chevron*,[22,25,26] *accordion*,[16,27] among others.[28–31] This plethora of geometrical models indicates the actual in- terest in developing and improving such a special and rare type of structure.

In this work, we propose a new type of topological model that presents near-zero PR under compression, which was experimentally realized using 3D printer technology. The model consists of an array of building blocks, as described below. It is important to stress that our model is not based on *semi-re-entrant*, *re-entrant* or regular honeycomb structures.

The building blocks of our model are composed of two parallel rigid bars with smooth surfaces and a soft and elastic membrane or spring connecting the rigid bars. Thereafter, the spring direction will be addressed as BAD - block axial direction. Fig. 1(c) presents the idealized model and Fig. 1(d) the 3D printed model where the curved membranes play the role of the spring.

The model array is generated by combining the building blocks in a way their BADs are mixed into two orthogonal directions, i. e., always parallel and/or perpendicular to each other. One example of a 2D array of 3 × 4 blocks is shown in Fig. 2. When compressed along the z direction by a $\sigma_z$ stress, the blocks that have their BADs parallel to z will get squeezed, following and accommodating $\varepsilon_z$ strain. The blocks with BADs orthogonal to the compression direction will only slide, without deformation, because of the rigid bars. The length of the system along x direction remains the same. These two mechanisms together led to zero PR behavior. If the stress is released, the system almost elastically recovers its initial configuration because the rigid bars have smooth surfaces and the friction energy dissipation can be neglected. Also, if we apply a compressive stress along the *x* direction instead of the *z* one, the effect will be the same, leading again to a zero PR. The mechanism of our model does not rely on a combination of positive and negative PR structures, as in the *semi-re-entrant* ones[22], but on the emergence of a zero PR in two dimensions from a system having already a zero PR in one direction. As the building blocks are not structurally bonded, it should be stressed that our model cannot be used in applications that depend on tensile stresses. In this sense, it can be considered as an intermediate structure between the fully connected structure of corks and the fully unconnected structure of the gases, both presenting zero or close to zero PR.[4] This model can be easily extended to a 3D system and with compression in any direction. It suffices to have building blocks with BAD oriented both parallel and perpendicular to each other along any direction.

Our model was 3D printed with a homemade system (fused deposition modeling)[32] using standard commercial polylactic acid thermoplastic filament. The soft spring is composed of two relatively thin membranes bonded at their middle and then to two rigid bars, Fig. 1(d). In Fig. 3, we present a 3 × 4 assembly of printed blocks used to demonstrate the null PR behavior.

A compressive stress was applied to the structure shown in Fig. 3, along the *z* direction, and the whole process was video recorded with a resolution of 1980 pixels × 1080 pixels and 30 fps (see the supplementary material). The *z* and *x* length values were obtained by processing, frame by frame, the experiment video using a python script. In Fig. 4 we present time evolution of the measured lengths along the z and x directions of the 3 × 4 assembled blocks during compression. It is clear from this Figure that, considering the error bars, the system size along x direction remains practically constant, thus characterizing a near-zero Poisson regime. From these data, we obtained PR = −0.003 ± 0.001.

These results show that our model provides an alternative way to create close to zero PR systems. Just for comparison, cork, beryllium, diamond and graphene sponge PRs are about 0.064,[13] 0.05,[33] ≤ 0.05[34] and 0.04,[35] respectively. Other relevant structures, such as SILICOMB proto- types made of ABS plastic have PR between −0.03 and 0.05,[23] patellar surface of canine femoral articular cartilage have PR ~ 0.07,[36] and semi re-entrant semiflexynes molecules obtained by molecular dynamics (MD) simulations present PR about 0.01.[25]

In summary, we propose a new type and very simple geometrical 2D model that presents near- zero PR under compression. It is important to stress that our model is not based on *semi-re-entrant*, *re-entrant* or regular honeycomb structures, which have been the basis of many structural zero PR reported in the literature. Our model was experimentally realized using 3D printer technology. The obtained PR values were PR = −0.003 ± 0.001, which are ones of the closest to zero PR values ever produced. This topological model can be easily extended to 3D systems and with compression along any direction. We hope the present work will stimulate further studies along these lines.

**SUPPLEMENTARY MATERIAL**

See supplementary material for the video showing the compression of our 3 × 4

system.

**ACKNOWLEDGMENTS**

A.F.F. and D.S.G. are fellows of the Brazilian Agency CNPq through grants numbers #311587/2018- 6 and #307331/2014-8, respectively. A.F.F. and D.S.G. acknowledge grants from São Paulo Research Foundation (FAPESP) numbers #2018/02992-4 and #2013/08293-7 respectively.

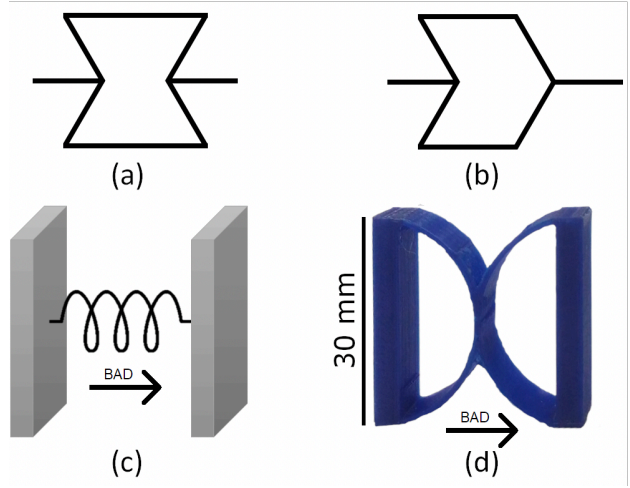

**FIG. 1.** Models for negative and zero PR structures. (a) re-entrant structure; (b) semi-re-entrant structure; (c) building block of the zero PR model proposed here, composed by two rigid bars (gray) and one soft spring connecting them; and (d) the 3D printing block where the curved membranes play the rule of springs in (c) model. The arrows below panels (c) and (d) indicate the corresponding direction of the *block axial direction* (BAD). See manuscript for details.

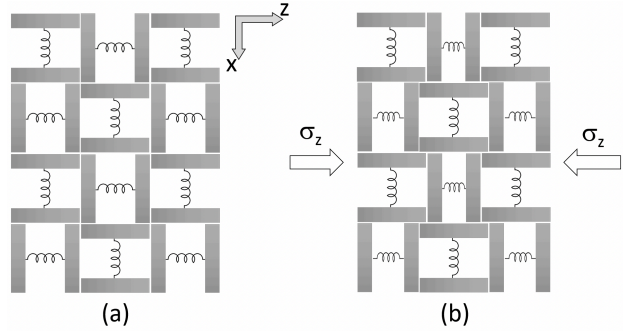

**FIG. 2.** (a) One example of an assembled zero PR system formed by an array of 3 × 4 blocks. (b) The expected structure after the action of a compressive stress, $\sigma_z$, applied along the $z$ direction.

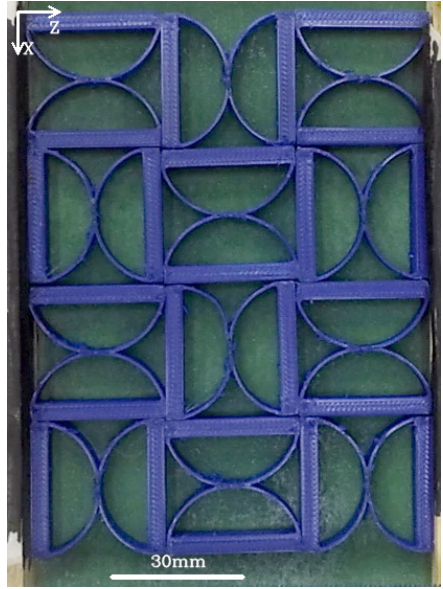

**FIG.** 3. 3 x 4 array of building blocks assembled with BAD's blocks oriented both along and perpendicular to a *z* direction. See the video in supplementary material for the compression of this structure.

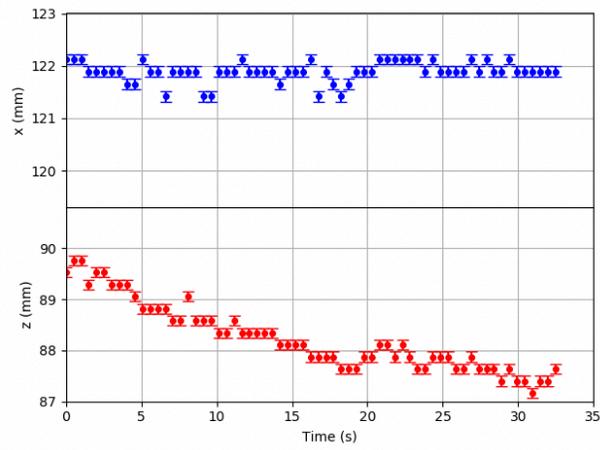

**FIG. 4.** Time evolution measurements of *z* and *x* lengths during compression in *z* direction.